\documentclass[aps,pra,floatfix,twocolumn,superscriptaddress,showpacs]{revtex4}
\usepackage{times}
\usepackage{amsmath}
\usepackage{bm}
\usepackage{dcolumn}                 
\newcolumntype{w}[1]{D{.}{.}{#1}}
\newcolumntype{.}{D{x}{}{-1}}

\newcommand{\balpha}{\vec{\alpha}}

\newcommand{\bfr}{\vec{r}}
\newcommand{\vare}{\varepsilon}
\newcommand{\lbr}{\langle}
\newcommand{\rbr}{\rangle}

\allowdisplaybreaks
\begin{document}
\title{Equation of motion for a bound system of charged particles}

\author{Krzysztof Pachucki}
\affiliation{
Faculty of Physics, University of Warsaw,
Pasteura 5, 02-093 Warsaw, Poland}

\author{Vladimir A. Yerokhin}
\affiliation{Center for Advanced Studies,
        Peter the Great St.~Petersburg Polytechnic University, Polytekhnicheskaya 29,
        St.~Petersburg 195251, Russia}

\begin{abstract}
We consider a bound system of charged particles
moving in an external electromagnetic field, including leading relativistic corrections.
The difference from the point particle with a magnetic moment comes
from the presence of polarizabilities.
Due to the lack of separation of the total momentum from the internal degrees of freedom,
the notion of polarizability of the bound state immersed in the continuum spectrum
of the  global motion is nontrivial.
We introduce a bound-continuum perturbation theory and obtain a complete formula for the equation of motion
for a polarizable bound system, such as atom, ion, or the nucleus.
This formula may find applications when high precision is sought and small
effects due polarizabilities are important.
\end{abstract}

\maketitle

\section{Introduction}
Let us consider a set of charged particles forming a bound system, like an atom, an ion, or a molecule.
We are interested in the motion and global properties
of this system in an external electromagnetic field. When relativistic corrections
are included, the center of mass cannot be separated from  internal degrees of freedom
and this causes the appearance of additional corrections to the electromagnetic moments and polarizabilities.
A typical problem is the magnetic moment of the bound system, for which the first complete
description was presented by Hegstrom in \cite{hegstrom}.
Here we rederive several results obtained in earlier works and
obtain the complete $1/M$ corrections to known polarizabilities.
Moreover, we derive polarizabilities that involve an additional field derivative and depend on a spin.
Results are summarized in terms of a general Hamiltonian Eq. (\ref{45}) and
equations of motion Eqs. (\ref{52},\ref{53}), which include all the leading relativistic corrections.
These results may find application when high precision is sought.
For example, as has already been noticed by Thompson {\em et al}. \cite{thompson},
the cyclotron frequency is slightly changed due to the electric dipole polarizability,
and this shift is actually measurable.
What has not yet been noticed is that the coupling of the spin to the static electromagnetic field
through spin dependent polarizabilities shifts the Larmor frequency, although
this effect is presently below the experimental accuracy \cite{gboron}.
Perhaps the most important application will be for calculations of nuclear polarizabilities,
which play a significant role in muonic atoms.

\section{Bound-continuum perturbative approach to separation of the center of mass motion}
In the first step we shall present a perturbative scheme for a bound system immersed
in the continuum spectrum of global motion,
when internal and external degrees of freedom are coupled together due to the presence
of an electromagnetic field. To the best of our knowledge, it was first introduced in Ref. \cite{wienczek},
and here we develop it further.
We assume that the Hamiltonian $H$ for a system of particles can be decomposed as
\begin{equation}
H =  H_{S} + H_{P}, \label{01}
\end{equation}
where $H_{S}$ is the Hamiltonian that involves only internal degrees of freedom
and does not depend on an external electromagnetic field.
$H_P$ is the remainder, in which all dependence on the external field is placed.

When an atom or an ion moves in the external electromagnetic
field, the internal degrees of freedom affect the center of mass
motion and the spin precession. We here assume that the coupling of
internal to global degrees of freedom is weak and does not affect
significantly the internal state and the motion. In this case, the
effects due to this coupling can be described perturbatively.
In the present work we formulate a systematic perturbative
approach to account for all these effects.
For this we assume that in the leading order our system is in the specified internal state $\psi_S$
with the binding energy $E_S$
\begin{align}
  H_{S}\,\psi_{S} =  { E}_{S}\,\psi_{S}\,. \label{02}
\end{align}
The initial wave function $\psi_\Pi$ describing the global motion of the system can be obtained as
\begin{align}
  \psi_\Pi = \langle\psi_{\rm S}|\psi\rangle_{S}\,, \label{03}
\end{align}
where $\langle\ldots\rangle_S$ denotes the matrix element in the subspace of internal degrees of freedom,
and hence $\psi_\Pi$  depends only on the global degrees of freedom.
Let us note, that the wave function $\psi_\Pi$ is redefined in Eq. (\ref{14})
to account for modifications of the internal state
due to external perturbations. The total wave function can thus be decomposed as
\begin{align}
  \psi =&\  \psi_{S}\,\psi_{\Pi} + \delta\psi\,. \label{04}
\end{align}
Let us introduce a projection operator $ P_\perp = I- |\psi_S\rangle\langle\psi_S|$;
then
\begin{align}
  P_\perp\psi =&\ \delta\psi\,. \label{05}
\end{align}
The Schr\"odinger equation for the whole system is
\begin{equation}
i \frac{\partial\psi}{\partial t}= (H_{\rm S} - E_{\rm S} + H_{\rm P})\,\psi, \label{06}
\end{equation}
where the binding energy $ E_{\rm S}$ was subtracted from the time dependence.
Let us project the Schr\"odinger equation (\ref{02}) into $\psi_{\rm S}$ and use Eq. (\ref{04})
\begin{equation}
\Bigl(i \frac{\partial}{\partial t} -\langle\psi_S|H_{\rm P}|\psi_S\rangle\Bigr)\,\psi_\Pi =
\langle \psi_{\rm S} | H_{\rm P} | \delta \psi \rangle_{\rm S}. \label{07}
\end{equation}
$\delta\psi$ in the right hand side is to be determined perturbatively as follows.
One projects the Schr\"odinger equation (\ref{06}) into the subspace orthogonal to $\psi_{\rm S}$ with $P_\perp$
  \begin{align}
    i \frac{\partial}{\partial t}\delta\psi =
     P_\perp\,(H_{\rm S} - E_{\rm S} +H_{\rm P})\,(\psi_{\rm S}\,\psi_\Pi + \delta\psi),\label{08}
  \end{align}
and solves the resulting equation, assuming $H_{\rm S} - E_{\rm S}$ is large
in comparison to $H_P$; then
    \begin{align}
      \delta\psi = \frac{1}{( E_{\rm S} - H_{\rm S})'}\,
      \Bigl[H_{\rm P}\,\psi_{\rm S}\,\psi_\Pi -\Bigl(i \frac{\partial}{\partial t}-H_{\rm P}\Bigr)\delta\psi\Bigr].\label{09}
    \end{align}
Another convenient form of the solution is
    \begin{align}
      \delta\psi = \frac{1}{\Bigl[( E_{\rm S} - H_{\rm S})'
          + P_\perp\,\bigl(i \frac{\partial}{\partial t} - H_{\rm P}\bigr)\,P_\perp\Bigr]}\,
       H_{\rm P}\,\psi_{\rm S}\,\psi_\Pi. \label{10}
    \end{align}
The replacement of Eq. (\ref{10}) into Eq. (\ref{07}) gives
\begin{widetext}
\begin{align}
\Biggl\{i \frac{\partial}{\partial t} -\langle\psi_S|H_{\rm P}|\psi_S\rangle
-\langle\psi_S|H_{\rm P}\,\frac{1}{\bigl[( E_{\rm S} - H_{\rm S})'
          + P_\perp\bigl( i \frac{\partial}{\partial t} - H_{\rm P}\bigr)\,P_\perp\bigr]}\,H_{\rm P}|\psi_S\rangle
  \Biggr\}\psi_\Pi = 0\,. \label{11}
\end{align}
Up to this point, our considerations have been exact, with no approximations involved.
We now introduce a perturbative expansion, assuming the
second term in the denominator to be small as compared to the first
one to obtain
\begin{align}
  &\Biggl\{i \frac{\partial}{\partial t} -\langle H_{\rm P}\rangle_S - \Biggl\langle\psi_S\Biggl|
  H_{\rm P}\frac{1}{( E_{\rm S} - H_{\rm S})'}H_{\rm P}
+H_{\rm P}\frac{1}{( E_{\rm S} - H_{\rm S})'}
\Bigl(H_{\rm P} - i \frac{\partial}{\partial t}\Bigr)
\frac{1}{( E_{\rm S} - H_{\rm S})'}H_{\rm P} + \cdots\Biggr|\psi_S\Biggr\rangle
  \Biggr\}\,\psi_\Pi = 0\,.\label{12}
\end{align}
The neglected terms denoted by dots include higher powers in the derivatives over time.
They are of the order of the ratio of the external field frequency to the excitation energy of the internal state,
which we assume to be small. Other neglected terms involve the ratio of the spatial field derivatives
over the mass $M$ and terms involving three powers of the external field, they are also assumed to be small.
The above equation, neglecting these higher-order terms,  is now transformed as follows:
\begin{align}
  &\Biggl\langle\psi_S\Biggl|H_{\rm P}\frac{\Bigl(i \frac{\partial}{\partial t}-\langle H_{\rm P}\rangle_S\Bigr)}
  {[( E_{\rm S} -H_{\rm S})']^2}H_{\rm P} \Biggr|\psi_S\Biggr\rangle =
\frac{1}{2}\Biggl\{\Biggl\langle\psi_S\Biggl|H_{\rm P}\frac{1}{[( E_{\rm S} -H_{\rm S})']^2}H_{\rm P}
\Biggr|\psi_S\Biggr\rangle\,,\,\Bigl(i \frac{\partial}{\partial t}- \langle H_{\rm P}\rangle_S\Bigr)\Biggr\} \nonumber \\ &\
+\frac{1}{2} \Biggl\langle\psi_S\Biggl|\biggl[H_{\rm P}, i \frac{\partial}{\partial t} - \langle H_{\rm P}\rangle_S\biggr]
\frac{1}{[( E_{\rm S} -H_{\rm S})']^2}H_{\rm P}\Biggr|\psi_S\Biggr\rangle
+\frac{1}{2} \Biggl\langle\psi_S\Biggl|H_{\rm P}\frac{1}{[( E_{\rm S} -H_{\rm S})']^2}
\biggl[i \frac{\partial}{\partial t} -\langle H_{\rm P}\rangle_S\,,\,H_{\rm P}\biggr]\Biggr|\psi_S\Biggr\rangle\,.\label{13}
\end{align}
The first term in the above can be eliminated by the redefinition  $\psi_\Pi$, namely
$\psi_\Pi = e^\lambda\,\tilde\psi_\Pi$, with
\begin{equation}
\lambda = -\frac{1}{2}\Biggl\langle\psi_S\Biggl|H_{\rm P}\frac{1}{[( E_{\rm S} -H_{\rm S})']^2}H_{\rm P}
\Biggr|\psi_S\Biggr\rangle\,.  \label{14}
\end{equation}
This brings the equation of motion to the more familiar form
\begin{align}
\Bigl(i \frac{\partial}{\partial t} - H_\Pi\Bigr)\tilde\psi_\Pi = 0\,, \label{15}
\end{align}
with
\begin{align}
  H_\Pi = &\
\Biggl\langle\psi_S\Biggl|H_{\rm P}
+H_{\rm P}\frac{1}{( E_{\rm S} - H_{\rm S})'}H_{\rm P}
+H_{\rm P}\frac{1}{( E_{\rm S} - H_{\rm S})'}\bigl(H_{\rm P} - \langle H_{\rm P}\rangle_S\bigr)
\frac{1}{( E_{\rm S} - H_{\rm S})'}H_{\rm P}\nonumber \\ &\
-\frac{1}{2}\biggl[H_{\rm P}\,,\,i \frac{\partial}{\partial t} - \langle H_{\rm P}\rangle_S\biggr]
\frac{1}{[( E_{\rm S} -H_{\rm S})']^2}H_{\rm P}
-\frac{1}{2}H_{\rm P}\frac{1}{[( E_{\rm S} -H_{\rm S})']^2}\biggl[i \frac{\partial}{\partial t} - \langle H_{\rm P}\rangle_S\,,
\,H_{\rm P}\biggr] \Biggr|\psi_S\Biggr\rangle\,. \label{16}
\end{align}
\end{widetext}
This is the general equation valid for a wide range of bound systems immersed into
slowly varying external fields, including nuclei. 
Let us now consider a simple case of the nonrelativistic bound system of charged particles
\begin{align}
 H_S =&\ \sum_a\frac{\vec q_a^{\,2}}{2\,m_a} +
\sum_{a>b,b}\frac{e_a\,e_b}{4\,\pi\,r_{ab}},\label{17} \\
H_{\rm P} =&\
\frac{\vec\Pi^2}{2\,M}+e\,A^0
- \vec d\cdot\vec E'\,, \label{18}
\end{align}
where $\vec d$ is the dipole moment operator, $\vec E'$ is the electric field as seen by the moving bound system
\begin{align}
  \vec E' =&\ \vec E + \Bigl(\frac{\vec\Pi}{M}\times\vec B\Bigr)_{\rm sym}\,, \label{19}
\end{align}
and where
\begin{align}
(\vec \Pi\times\vec B)^i_{\rm sym} =&\ \frac{1}{2}\,\epsilon^{ijk}\,\{\Pi^j\,,\,B^k\}\,. \label{20}
\end{align}
With these nonrelativistic $H_S$ and $H_P$ Hamiltonians,  $H_\Pi$  takes the following form:
\begin{widetext}
\begin{align}
  H_\Pi = &\
  \frac{\vec\Pi^2}{2\,M}+e\,A^0 + E'^i\,E'^j\,
  \Biggl\langle\psi_S\Biggl|d^i\frac{1}{( E_{\rm S} - H_{\rm S})'}\,d^j\Biggr|\psi_S\Biggr\rangle
  \nonumber \\&\
 +\frac{1}{2}\,
 \biggl\{\biggl[E^i\,,\, -i \frac{\partial}{\partial t} + \frac{\Pi^2}{2\,M}\biggr]\,E^j
 +E^i\,\biggl[-i \frac{\partial}{\partial t} + \frac{\Pi^2}{2\,M}\,,\,E^j\biggr] \biggr\}
 \Biggl\langle\psi_S\Biggl|d^i\frac{1}{[( E_{\rm S} - H_{\rm S})']^2}\,d^j\Biggr|\psi_S\Biggr\rangle \label{21} \\ = &\
  \frac{\vec\Pi^2}{2\,M}+e\,A^0 +
  \frac{1}{2}\,\biggl[\{E'^i\,,\,E'^j\} + \frac{i\,\epsilon^{ijk}}{M}\,B^l\,E^l_{,k} \biggr]
  \,\Biggl\langle\psi_S\Biggl|d^i\frac{1}{ E_{\rm S} - H_{\rm S}}\,d^j\Biggr|\psi_S\Biggr\rangle
  \nonumber \\&\
+\frac{1}{2}\,
\biggl[ \frac{1}{M}\,E^i_{,k}\,E^j_{,k} + i\,(\dot E^i\,E^j -E^i\,\dot E^j) + \frac{1}{2\,M}\,(E^i \Pi^2 E^j - E^j \Pi^2 E^i)\biggr]
\Biggl\langle\psi_S\Biggl|d^i\frac{1}{[ E_{\rm S} - H_{\rm S}]^2}\,d^j\Biggr|\psi_S\Biggr\rangle\,. \label{22}
\end{align}
The antisymmetric in the $i,j$ part corresponds to the vector polarizability and leads to an additional spin precession.

In the second example we add to $H_{\rm P}$ the following term
\begin{equation}
  \delta H_{\rm P} = i\,\vec\Pi\,[\vec t,H_S]\,, \label{23}
\end{equation}
with some arbitrary Hermitian operator $\vec t$. Such a term appears after separation of the center-of-mass motion
for the bound system of relativistic particles \cite{compound}, and we demonstrate here how to deal with it.
The resulting correction to $H_\Pi$ is
\begin{align}
  \delta H_\Pi =&\ \biggl\langle\psi_S\biggl|H_{\rm P}\frac{1}{( E_{\rm S} - H_{\rm S})'}\delta H_{\rm P}
+H_{\rm P}\frac{1}{[( E_{\rm S} - H_{\rm S})']^2}\biggl[e\,A^0+\frac{\Pi^2}{2\,M} - i\frac{\partial}{\partial t}\, ,\,
\delta H_{\rm P}\biggr]\biggr|\psi_S\biggr\rangle
\nonumber \\ &\ +
 \biggl\langle\psi_S\biggl|\delta H_{\rm P}\frac{1}{( E_{\rm S} - H_{\rm S})'}H_{\rm P}
+\biggl[\delta H_{\rm P}\, , \, e\,A^0+\frac{\Pi^2}{2\,M} - i\frac{\partial}{\partial t}\biggr]
\frac{1}{[( E_{\rm S} - H_{\rm S})']^2} H_{\rm P}\biggr|\psi_S\biggr\rangle\,. \label{24}
  \end{align}
The commutator in the above is
\begin{align}
  \biggl[e\,A^0+\frac{\Pi^2}{2\,M} - i\frac{\partial}{\partial t}\, ,\, i\,\vec t\,\vec \Pi\biggr] =&\
  \biggl(\vec E+\frac{\vec \Pi}{M}\times\vec B\biggr)_{\rm sym}\,\vec t = \vec E'\,\vec t\,, \label{25}
\end{align}
and $\delta H_\Pi$ becomes
\begin{align}
  \delta H_\Pi =&\ \bigl\langle i\bigl[\vec d\,\vec E'\,,\,\vec t\,\vec \Pi\bigr]\bigr\rangle +
  \biggl\langle\psi_S\biggl|\vec d\,\vec E'\frac{1}{( E_{\rm S} - H_{\rm S})'}\vec t\,\vec E'\biggr|\psi_S\biggr\rangle
+ \biggl\langle\psi_S\biggl|\vec t\,\vec E'\frac{1}{( E_{\rm S} - H_{\rm S})'}\vec d\,\vec E'\biggr|\psi_S\biggr\rangle\,. \label{26}
\end{align}
\end{widetext}
Alternatively, this result can be obtained by a unitary transformation
of the original Hamiltonian, as follows
\begin{align}
  H' = e^{-i\,\phi}\,H\, e^{i\,\phi} +\partial_t\phi = H + i\,[H-i\,\partial_t\,,\,\phi]\,, \label{27}
\end{align}
with
\begin{align}
\phi = -\vec t\,\vec\Pi\,. \label{28}
\end{align}
The resulting new Hamiltonian
\begin{align}
H' = H_S -{E}_S + H_P - \vec t\,\vec E' + i\bigl[\vec d\,\vec E'\,,\,\vec t\,\vec \Pi\bigr] \label{29}
\end{align}
leads to the same $H_\Pi$, although in a much more straightforward  way.
This demonstrates the internal consistency of the bound-continuum perturbative formalism.
In the next section we apply it to the most general Hamiltonian of a bound system
and derive formulas for polarizabilities.

\section{Hamiltonian for a bound system interacting with an electromagnetic field}
Let us assume that the electric and magnetic potentials have linear and quadratic components,
and the field strengths are time independent. The general Hamiltonian was derived from
the first principles in Refs. \cite{compound, wienczek} by a sequence of unitary transformations
of $H = \sum_a H_a + \sum_{a>b} H_{ab}$, where $H_a$ is a one particle Hamiltonian with relativistic corrections
and $H_{ab}$ is a two-body Hamiltonian with Coulomb interactions and relativistic corrections,
\begin{widetext}
\begin{align}
  H =&\ H_S + \frac{\vec\Pi^2}{2\,M}\,\biggl(1-\frac{H_S}{M}\biggr) - \frac{\Pi^4}{8\,M^3}
  + e\,A^0 -(\vec d + \delta\vec d)\cdot\vec E
  -\vec d\cdot\biggl(\frac{\vec \Pi}{M}\times \vec B\biggr)_{\rm sym}
  -\frac{1}{2}\,(d^{ij} + \delta d^{ij})\,E^i_{,j}
  -\frac{1}{2}\,\mu^{ij}\,B^j_{,i}
  \nonumber \\ &\
  - (\vec\mu+\delta{}\vec\mu)\cdot\vec B
+ \biggl(\vec\mu
-\frac{e}{2\,M}\,\vec S\biggr)\biggl(\frac{\vec\Pi}{M}\times\vec E\biggr)_{\rm sym}
+ \frac{3}{8\,M}\,(\vec d\times\vec B)^2
+\frac{1}{8}\,B^i\,B^j\,(\delta^{ij}\,d'^{kk}-d'^{ij})
 - \frac{e}{2\,M}\,E^i\,E^j\,d^{ij}\,, \label{30}
\end{align}
\end{widetext}
where $E^i_{,j} = \partial E^i/\partial R^j$, and we have introduced the following global variables:
the center of mass $\vec R$ and the total momentum $\vec \Pi$
\begin{align}
\vec R =&\ \sum_a \frac{m_a}{M}\,\vec r_a\,, \label{31}\\
\vec \Pi =&\ \sum_a \bigl[\vec p_a-e_a\,\vec A(\vec R)\bigr] = \vec P-e\,\vec A(\vec R)\,,
\label{32}
\end{align}
with $M = \sum_a m_a$ and $e = \sum_a e_a$, and relative coordinates
\begin{align}
\vec x_a =&\ \vec r_a-\vec R\,, \label{33}\\
\vec q_a =&\ \vec p_a-\frac{m_a}{M}\,\vec P\,, \label{34}
\end{align}
such that
\begin{align}
\bigl[x_a^i\,,\,q_b^j\bigr] &=\
i\,\delta^{ij}\,\biggl(\delta_{ab}-\frac{m_b}{M}\biggr)\,, \label{35}\\
\bigl[R^i\,,\,P^j\bigr] &=\  i\,\delta^{ij}\,,  \label{36}\\
\bigl[x_a^i\,,\,P^j\bigr] &=\  \bigl[R^i\,,\,q_a^j\bigr] = 0\,. \label{37}
\end{align}
The above Hamiltonian includes the following electromagnetic moments
\begin{align}
  \mu^i =&\ \sum_a \frac{e_a}{2\,m_a}\,(l_a^i + g_a\,s_a^i)\,, \label{38} \\
  \mu^{ij} =&\ \sum_a \frac{e_a}{m_a}\,\Bigl[g_a\,x_a^i\,s_a^i + \frac{1}{3}\,(l_a^j\,x_a^i + x_a^i\,l_a^j)\Bigr]\,, \label{39} \\
  S^i =&\ \sum_a {l}_a^i +{s}_a^i\,, \label{40} \\
  d^i =&\ \sum_a e_a\,x_a^i\,, \label{41} \\
  d^{ij} =&\ \sum_a e_a\,x_a^i\,x_a^j\,, \label{42} \\
  d'^{ij} =&\ \sum_a \frac{e_a^2}{m_a} x_a^i\,x_a^j\,, \label{43}
\end{align}
and relativistic corrections to them \cite{wienczek}
\begin{align}
  \delta\vec d =&\ -\frac{e}{2M} \sum_a\biggl(\frac{ q^j_a\,\vec x_a\,q^j_a}{m_a}
  +\sum_{b \neq a} \frac{e_a\,e_b}{4\,\pi} \frac{\vec x_a}{r_{ab}}\biggr)\,, \label{44}
\end{align}
$\delta\vec\mu$ is given in Refs. \cite{hegstrom, wienczek}
and we neglect  $\delta d^{ij}$ the relativistic correction to the electric quadrupole operator.
Moreover, the electromagnetic fields in Eq. (\ref{30}) and below are at the mass center $\vec R$.

Using the bound-continuum perturbative approach, namely  Eq. (\ref{16}),
$H$ in eq. (\ref{30}) gives the following effective Hamiltonian $H_\Pi$ for the motion
in the external static electromagnetic field
\begin{widetext}
\begin{align}
  H_{\Pi} =&\
  e\,A^0
+ \frac{\Pi^2}{2\,M}
-\frac{\Pi^4}{8\,M^3}
-\frac{e\,g}{2\,M}\vec S\cdot \vec B
+ \frac{e\,(g-1)}{2\,M^2}\vec S\cdot\vec\Pi\times\vec E
-\frac{Q}{6}\,(S^i\,S^j)^{(2)}\,E^i_{,j}
\nonumber \\ &\
-\frac{\alpha_E^{ij}}{4}\,\{E'^i\,,\,E'^j\} - \frac{\alpha_M^{ij}}{2}\,B^i\,B^j
- \frac{\alpha^{ijk}_{V\!M}}{4}\,B^i\,(E^j_{,k} + E^k_{,j})
- \frac{\alpha^{ijk}_{V\!E}}{4}\,E^i\,(B^j_{,k} + B^k_{,j})\,, \label{45}
\end{align}
\end{widetext}
where the mass $M$ of the system includes now the total binding energy,
$g$ factor is defined by
\begin{equation}
\bigl\langle \vec\mu\bigr\rangle\equiv
\frac{e}{2\,M}\,g\,\vec S,  \label{46}
\end{equation}
the quadrupole moment $Q$ is
\begin{align}
  \bigl\langle 3\,d^{ij}- d^{kk}\,\delta^{ij}\bigr\rangle =  Q\,(S^i\,S^j)^{(2)}\,, \label{47}
\end{align}
and where
\begin{align}
  \alpha_E^{ij} =&\ 2\, \Biggl\langle(d^i+\delta d^i)
  \frac{1}{ H_{\rm S} - E_{\rm S}}\,(d^j+\delta d^j)\Biggr\rangle +\frac{e}{M}\,\langle d^{ij}\rangle\,,  \label{48}\\
  \alpha_M^{ij} =&\ 2\, \Biggl\langle\mu^i\frac{1}{H_{\rm S} - E_{\rm S}}\,\mu^j\Biggr\rangle
  -\frac{3}{4\,M}\langle \vec d^{\,2}\,\delta^{ij}-d^i\,d^j\rangle\,, \label{49}
\nonumber \\ &\
  -\frac{1}{4}\,\langle d'^{kk}\,\delta^{ij} - d'^{ij}\rangle \\
  \alpha^{ijk}_{V\!M} =&\ -i\,\frac{\delta^{ij}}{M}\epsilon^{mnk} \Biggl\langle d^m\frac{1}{H_{\rm S} - E_{\rm S}}\,d^n \Biggr\rangle
\nonumber \\ &\
  +\Biggl\langle \mu^i\frac{1}{H_{\rm S} - E_{\rm S}}\,d^{jk}  + {\rm h.c.}\Biggr\rangle\,,  \label{50}\\
  \alpha^{ijk}_{V\!E} =&\ \biggl\langle d^i\, \frac{1}{H_{\rm S} - E_{\rm S}}\,\mu^{jk} + {\rm h.c.}\biggr\rangle\,,   \label{51}
\end{align}
and we neglect the electric quadrupole polarizability.
The first part of the Hamiltonian in Eq. (\ref{45}) is very much unique,
the coupling of spin to the electromagnetic field
is fully described by a BMT equation \cite{bmt}, and the result obtained here is in agreement
with the small momentum expansion of this equation.
The second part  of $H_\Pi$ involves intrinsic polarizabilities, which solely depend on the internal Hamiltonian $H_S$.

\section{Applications}
The resulting equations of motion  are
\begin{align}
  \frac{d\,\vec \Pi}{d t} =&\   i\,\bigl[H_\Pi\,,\,\vec\Pi\bigr] + \frac{\partial \vec\Pi}{\partial t}\,,  \label{52}\\
  \frac{d\vec S}{d t} =&\  i\,\bigl[H_\Pi\,,\,\vec S\bigr]\,. \label{53}
\end{align}
The first observation is the presence of the spin-dependent force in the homogeneous electromagnetic field,
namely
\begin{align}
  \vec F = -\frac{e^2\,(g-1)}{2\,M^2}\,\vec B\times(\vec E\times\vec S)\,. \label{54}
\end{align}
Being suppressed by $\mu\,B/M$ with respect to the Lorentz force,
it is a very small force and we are not aware of any experiment that demonstrates its presence.

The  case of a charged particle in a Penning trap has been analyzed in detail in
Ref. \cite{gabrielse}, but without polarizabilities.
The interesting effect due to the scalar electric dipole polarizability,
which has already been observed \cite{thompson}, is the shift of the cyclotron frequency $\omega_C$.
Let us assume a nonrelativistic spinless system, for which the Hamiltonian is
\begin{align}
  H_{\Pi} =&\ e\,A^0 + \frac{\Pi^2}{2\,M} - \frac{\alpha_E}{2}\,\vec E'\cdot\vec E'\,.  \label{55}
\end{align}
In a classical picture the ion stays in the region where $\vec E$ vanishes in the minimum
of the electromagnetic (axially symmetric) potential $A^0$, so
\begin{align}
  H_{\Pi} =&\ \frac{\Pi^2}{2\,M} -\frac{\alpha_E}{2\,M^2}\,(\vec\Pi\times\vec B)^2 \nonumber\\
  =&\, \frac{P_z^2}{2\,M} + \frac{\Pi_x^2+\Pi_y^2}{2\,M} \biggl(1-\frac{\alpha_E\,B^2}{2\,M^2}\biggr)\nonumber\\
  =&\, \frac{P_z^2}{2\,M} + \frac{\Pi_x^2+\Pi_y^2}{2\,M_\perp}\,, \label{56}
\end{align}
where $M_\perp = M+\alpha_E\,B^2/(2\,M)$. As a result, the cyclotron frequency $\omega_C = e\,B/M_\perp$
is slightly shifted by the polarizability $\alpha_E$ \cite{thompson}.

The magnetic polarizability is usually calculated neglecting $1/M$ corrections because it is very small.
Nonetheless we will derive the closed formula for the leading recoil correction,
because it affects the accurate determination of the refractive index of He
\cite{puchalski:16}. Our result for the scalar magnetic polarizability
\begin{align}
  \alpha_M =&\ \frac{2}{3}\,\Biggl\langle\mu^i\frac{1}{H - E}\,\mu^i\Biggr\rangle
  -\frac{\langle d^i\,d^i\rangle}{2\,M}
  -\frac{\langle d'^{ii}\rangle}{6} \label{57}
\end{align}
includes $1/M$ effects exactly. After expanding this formula in the electron-nuclear mass ratio,
one obtains
\begin{align}
  \alpha_M =&\ -\frac{e^2}{6\,m}\,\Big\langle\sum_a \vec r_a^{\,2}\Big\rangle
  -\frac{e^2}{6\,M}\,\sum_{a,b}\langle \vec r_a\,\vec r_b\rangle + O\Big(\frac{m}{M}\Big)^2 \,, \label{58}
\end{align}
in agreement with Ref. \cite{bruch} for the particular case of the helium atom.

Coming back to the Penning trap, the vector polarizability $\alpha_{V\!M}$ in Eq. (\ref{50})
depends on the spin, and thus contributes to the Larmor frequency. This frequency is accurately
measured in the determination of the bound electron $g$ factor, where ion is placed \cite{blaum} in
the homogeneous magnetic field $\vec B = (0,0,B)$ and the quadrupole electric field
\begin{align}
  e\,A^0 =&\ U_0\,\biggl(z^2-\frac{x^2+y^2}{2}\biggr)\,,\label{59}
\end{align}
where $e<0$ is the electron charge. For an ion with a spinless nucleus the corresponding shift
of an energy level is
\begin{align}
\delta E_{V\!M} = \biggl\langle \vec\mu\cdot\vec B\,\frac{1}{E-H}\,\frac{1}{2}\,d^{ij}\,E^i_{,j}\biggr\rangle + {\rm h.c.}
  \label{60}
\end{align}

This correction may get an enhancement if there is an excited state with a small energy difference
from the reference state. This is the case for boronlike ions, where the first excited $2P_{3/2}$
state is close to the ground  $2P_{1/2}$ state, specifically, for boronlike argon Ar$^{13+}$
measured in Ref.~\cite{gboron} at the $10^{-9}$ accuracy level.

We calculate this effect as described in the Appendix, representing the corresponding correction to the
$g$ factor as
\begin{align}
\delta g_{V\!M} = 2\,U_0\,{\cal R}\,, \label{61}
\end{align}
where $U_0$ is the parameter of the quadrupole electric field in the trap  (\ref{59}) and ${\cal
R}$ is the atomic part. The parameter $U_0$ can be expressed in terms of the measured axial frequency
$\nu_z$ of the ion in the trap
\begin{align}
  U_0 = \frac{1}{2\,Q}\,\frac{M}{m}\,\frac{(2\,\pi\,\hbar\,\nu_z)^2}{(m\,c^2)^2}\,, \label{62}
\end{align}
where  $Q$ is the charge number of the ion, $M$ is the ion mass, and $m$ is the electron mass.

The obtained result for the ground 2P$_{1/2}$ state of $^{40}$Ar$^{13+}$ is
\begin{align}
  \delta g_{V\!M} = 1.7\cdot 10^{-17}\,,\label{63}
\end{align}
which is much below the present experimental accuracy \cite{gboron}. It is interesting that the
effect is strongly $Z$-dependent and increases for smaller $Z$. For example, for boron-like carbon
it is about four orders of magnitude larger (see the Appendix) but still too small to be of the
experimental interest at present.

The last example is the relativistic correction to the electric dipole polarizability for
a system with nonvanishing charge. The neutral case  $e=0$ has been investigated in detail,
in particular for the helium atom \cite{puchalski:16}, for which all relativistic corrections
to the electric dipole polarizability come from the internal Hamiltonian, and there are no corrections
to the electric dipole moment operator. However, in the case $e\neq 0$ there are
additional corrections, specifically
\begin{align}
  \delta\alpha_{E} = \frac{4}{3}\,\biggl\langle \vec d
  \frac{1}{H - E}\,\delta \vec d\biggr\rangle
  + \frac{e}{3\,M}\,\big\langle \sum_a e_a x_a^2\big\rangle\,,\label{64}
\end{align}
where $\vec d$ is the electric dipole operator with respect to the mass center,
$\delta\vec d$ is defined in Eq. (\ref{44}),
and the last term can be interpreted as the mean square charge radius.
$\delta\alpha_{E}$ is a small correction in the case of ions, but it can be significant in the case of nuclei.
Because this formula has been obtained for purely electromagnetic systems, it may not be valid
for a nucleus, but it indicates the existence of corrections to polarizabilities,
which come from the separation of internal and global degrees of freedom. Such corrections
may play a role in muonic atoms, where significant discrepancies have been observed \cite{aldo},
particularly for the hyperfine splitting \cite{fs_hfs}.

\section{Summary}
In this work we have formulated a perturbation theory that allows the systematic derivation of
the properties of a bound charged system in the external electromagnetic field. It is found that
for slowly changing electromagnetic fields, the perturbation theory provides the equation of motion
for a point particle with magnetic dipole and electric quadrupole moments, and additionally includes
different types of polarizabilities. It is an obvious fact, but it has not yet been shown
with including relativistic corrections.
We have derived formulas for dipole polarizabilities including finite nuclear mass effects
and new types of polarizabilities, which include one more derivative in the electromagnetic field.
Several examples demonstrate the potential applications of the derived formulas.

\begin{acknowledgments}
  This work was supported by the National Science Center (Poland) Grant No. 2017/27/B/ST2/02459.
  V.A.Y. acknowledges support by the Ministry of Education and Science
  of the Russian Federation Grant No. 3.5397.2017/6.7.
\end{acknowledgments}

\appendix
\section{Correction to the atomic $g$ factor due to quadrupole electric field}

In this section, we perform a relativistic calculation of the correction to the $g$ factor induced
by the quadrupole electric field in the Penning trap. Within the independent-electron
approximation, the leading contribution to the $g$ factor of an alkali-metal-like atom (one electron $v$
beyond the closed shells) is given by
\begin{align}
g &\ = \frac1{\mu_0\,B_z\,\mu_v}\,\delta E
 \nonumber \\ &
 =  \frac1{\mu_0\,B_z\,\mu_v}\,\langle v | (-e/2)\,B_z\, [\bfr \times \balpha]_z |v\rangle
 \nonumber \\ &
 =  \langle v | \frac{1}{\mu_v}\,[\bfr \times \balpha]_z |v\rangle\,,
\end{align}
where $\mu_0 = |e|/2$ is the Bohr magneton, $e<0$ is the electric charge, $B$ is assumed to be
directed along the $z$ axis, and $\mu_v$ is the momentum projection of the valence state $v$.

The electrostatic quadrupole potential in the Penning trap is \cite{blaum}
\begin{align}
e\,\Phi(r) = U_0\,\Big( z^2 - \frac{x^2+y^2}{2}\Big) = U_0\,\sqrt{\frac{4\pi}{5}}\,r^2\,Y_{20}(\bfr)\,,
\end{align}
where $Y_{lm}$ is the spherical harmonics. The correction to the $g$ factor due to $\Phi$ appears
as a second-order perturbative correction. Fixing the momentum projection of the valence state as
$\mu_v = 1/2$, we obtain
\begin{align}
\delta g = &\ 2\,U_0 \sum_{n\ne v} \frac1{\vare_v-\vare_n}
 \lbr v\,1/2|  2\,[\bfr \times \balpha]_z |n\,1/2\rbr
 \nonumber \\ &
 \times
 \lbr n\,1/2| \sqrt{\frac{4\pi}{5}}\, r ^2\, Y_{20}(\bfr)|v\,1/2\rbr \,.
\end{align}
This correction is enhanced for B-like ions because of large mixing between the $2p_{1/2}$ and
$2p_{3/2}$ states. In this case, we keep only the term with the smallest denominator in the sum
over $n$, obtaining
\begin{align}
\delta g = &\ 2\,U_0 \frac1{\vare_v-\vare_{v'}}
 \lbr v\,1/2|  2\,[\bfr \times \balpha]_z |v'\,1/2\rbr
 \nonumber \\ &
 \times
 \lbr v'\,1/2| \sqrt{\frac{4\pi}{5}}\, r ^2\, Y_{20}(\bfr)|v\,1/2\rbr \,,
\end{align}
where $v = 2p_{1/2}$ and $v' = 2p_{3/2}$. Performing angular integrations, we get
\begin{align}
{\cal R} \equiv & \lbr v\,1/2|  2\,[\bfr \times \balpha]_z |v'\,1/2\rbr\,
 \lbr v'\,1/2| \sqrt{\frac{4\pi}{5}}\, r ^2\, Y_{20}(\bfr)|v\,1/2\rbr
 \nonumber \\ &
= -\frac{4}{15}\,\Big[\int_0^{\infty}dr\,r^3\,(g_vf_{v'}+f_vg_{v'})\Big]
 \nonumber \\ &
 \ \ \ \ \ \ \ \ \ \ \ \ \ \ \ \ \ \times
\Big[\int_0^{\infty}dr\,r^4\,(g_vg_{v'}+f_vf_{v'})\Big]
 \,.
\end{align}
With the hydrogenic wave functions, we obtain
\begin{align}
{\cal R}  = \frac{8}{(Z\alpha)^2} - \frac{47}{10} + O((Z\alpha)^2)\,,
\end{align}
and
\begin{align}\label{eq001}
\delta g = 2\,U_0\,\Big[ -\frac{256}{(Z\alpha)^6} + \frac{1552}{5(Z\alpha)^4}+ \ldots \Big]\,.
\end{align}

We present more accurate results for two specific cases of B-like argon and carbon. Calculating
radial integrals with wave functions obtained with the localized Dirac-Fock potential, we get
${\cal R}  = 615 \pm 8\%$ for argon and ${\cal R}  = 1.3\times10^4 \pm 60\%$ for carbon, where the uncertainty
estimates the residual electron-correlation effects. Using the experimental values of the
$2p_{3/2}$\,--\,$2p_{1/2}$ energy splitting ($0.206\,458$~Ry for argon and $0.000\,578$~Ry for
carbon), we arrive at (in relativistic units)
\begin{align}
\delta g({\rm Ar}^{13+}) = 2\,U_0\,\Big[ -1.12 \times 10^8 \Big]\pm 8\% \,,\\
\delta g({\rm C}^{+})  = 2\,U_0\,\Big[ -8.7 \times 10^{11} \Big]\pm 60\%
\,.
\end{align}

The parameter $U_0$ characterizes the trapping potential and is specific for each experiment. We
obtain the value of $U_0$ in the experiment on B-like argon \cite{blaum} from the reported value of
the axial frequency $\nu_z$ of the ion in the trap, as
\begin{align}
  U_0 = \frac{M}{2\,Q}\,(2\,\pi\,\nu_z)^2\,,
\end{align}
where $Q$ is the charge number of the ion. Since all the calculations have been performed with
the electron mass $m=1$, it should be converted to dimensionless units as follows
\begin{align}
  U_0 = \frac{1}{2\,Q}\,\frac{M}{m}\,\frac{(2\,\pi\,\hbar\,\nu_z)^2}{(m\,c^2)^2}\,.
\end{align}
Using physical constants
\begin{align}
  \hbar =&\ 6.582\times 10^{-16}\; {\rm eV\, s}\,,\\
  m\,c^2 =&\ 0.511\times 10^6\; {\rm eV}\,,
\end{align}
parameters of the ion $Q = 13$, $  M/m = 72833.97$, and the measured axial frequency \cite{gboron}
$\nu_z = 650$~kHz, we arrive at
\begin{align}
  U_0(^{40}{\rm Ar}^{13+}) = 7.752 \times 10^{-26}\,.
\end{align}

\end{document}